\documentstyle[12pt,bezier]{article}
\newfont{\mathea}{msam10 scaled\magstep0}
\newfont{\matheb}{msbm10 scaled 1095}
\newfont{\tmpEins}{cmsy10 scaled 2074}
\newfont{\tmpZwei}{cmsy10 scaled 1095}
\newfont{\tmpDrei}{cmsy10 scaled 1000}
\newfont{\tmpVier}{cmsy5 scaled 1000}
\newfont{\tmpFuenf}{msbm7 scaled\magstep0}

\def\dach#1#2{\mbox{$\mathop{\vbox{\ialign{%
  $##\crcr\hfil #1 \hfil$\crcr}}}\limits^{\scriptscriptstyle #2}$}}

\def\rnzs{\dach{\rho_2}{\mbox{$\scriptscriptstyle\kern-.7mm0$}}\kern-1.2mm'}

\def\Subset{\mbox{$\subset\kern-.5mm\subset$}}
\newcommand{\LI}{\mbox{{\rm L$^{\kern-.15em\raise.2ex\hbox{\scriptsize 1}}$}}}

\def\Ldummy{\left.\bgroup}
\def\Rdummy{\egroup^{\rule{0mm}{1.4mm}}\right.}
\def\LA{\left\langle\bgroup}
\def\RA{\egroup^{\rule{0mm}{1.4mm}}\right\rangle_{\cal A}^{}}
\def\LR{\left(\bgroup}
\def\RR{\egroup^{\rule{0mm}{1.4mm}}\right)}
\def\LG{\left\{\bgroup}
\def\RG{\egroup^{\rule{0mm}{1.4mm}}\right\}}

\def\Wort#1{\mbox{{\rm #1\kern.1em}}}

\def\lfac#1#2{\vcenter{\hbox{$#1\kern-.2em\raise-.6ex\hbox{\Large{/}}%
 \kern-.2em\raise-1.2ex\hbox{$#2$}$}}}

\def\gin{\mbox{\tmpZwei\symbol{91}\kern-1.4mm\rule{.2mm}{1.85mm}\kern1.4mm}}
\def\gni{\mbox{\tmpZwei\symbol{92}\kern-1.4mm\rule[.15mm]{.2mm}{1.85mm}%
  \kern1.4mm}}

\def\EINS{{\mathop{1\kern-.25em\mbox{{\rm{\small l}}}}}}

\begin{document}
\Large The general Friedmann equation: a mathematical point of view

\vspace{3mm}

\normalsize Hellmut Baumg\"artel

Mathematical Institute

University of Potsdam

Germany

e-mail: baumg@uni-potsdam.de

\begin{abstract}

The note presents a classification of the relevant distinct types of solutions of the 
general Friedmann equation without assuming a priori restrictions for the parameters
occurring in this equation. The emphasis is on the case of a non-vanishing cosmological 
constant. The classification uses algebraic criteria. The result is: There are four
distinct basic types of models. Explicit formulas for decisive terms are given. 
Characteristic mutual relations of cosmological constant, mass and radiation density
to distinguish between the models are calculated.

\end{abstract}

\section{Introduction}

The Friedmann equation is a model for the description of a universe satisfying Einsteins
field equations of the general relativity theory and Einsteins cocmological principle,
expressed by the ansatz of a Robertson-Walker metric. The present paper describes the 
distinct types of solutions of this equation, using algebraic criteria for the classification.
Explicit formulas for decisive terms are presented. We do not assume a priori restrictions
for the parameters occurring in this equation.

For convenience a brief outline of the derivation of the Friedmann equation from the
mentioned starting point is given (see e.g. [1]). Einsteins field equations read
\begin{equation}
(R_{\mu,\nu}-\frac{1}{2}g_{\mu,\nu}R)+\Lambda g_{\mu,\nu}+\kappa T_{\mu,\nu}=0,\quad 
\kappa :=\frac{8\pi G}{c^{2}},
\end{equation}
where $\{g_{\mu,\nu}\}$ denotes the metric tensor, $\{R_{\mu,\nu}\}$ the "Verj\"ungung"
of Riemanns curvature tensor, $R:=g^{\mu,\nu}R_{\mu,\nu}$, where the sum convention is 
used, $\{T_{\mu,\nu}\}$ the energy tensor and $\Lambda$ the cosmological constant. $G$ is
the gravitational constant and $c$ the velocity of light. Further Einsteins cosmological
principle says: The universe is homogeneous and isotropic w.r.t. space. The ansatz
\begin{equation}
ds^{2}=c^{2}dt^{2}-R(t)^{2}\left(\frac{dr^{2}}{1-\epsilon r^{2}}+r^{2}(d\theta^{2}
+\sin^{2}\theta d\phi^{2})\right)
\end{equation}
of Robertson-Walker for the metric tensor corresponds to this principle. $r,\theta,\phi$
are dimension-less and time-independent coordinates. The dynamics is determined by the
scale-factor $R(t)$ only. It has the dimension of a length. $\epsilon$ is the curvature
parameter with values $\epsilon=0,+1,-1$ corresponding to the euclidean, spherical and
hyperbolic case, respectively. The energy tensor is given by
\begin{equation}
T_{\mu,\nu}:=\left(\rho+\frac{P}{c^{2}}\right)u_{\mu}u_{\nu}-g_{\mu,\nu}P,
\end{equation}
where $\rho,P,u_{\mu}$ denote mass density, pressure and 4-velocity, respectively. The
cosmological principle requires that $\rho$ and $P$ are homogeneous w.r.t. space, i.e.
\[
\rho=\rho(t),\quad P=P(t),\quad u=\{c,0,0,0\}
\]
where $t$ is the eigen-time. According to (2) and (3) one obtains
\begin{equation}
\{T_{\mu,\nu}\}=\mbox{diag}\,\left\{\rho c^{2},\frac{PR^{2}}{1-\epsilon r^{2}},
PR^{2}r^{2},PR^{2}r^{2}sin^{2}\theta\right\}.
\end{equation}
Putting (4) and (2) into (1) one obtains for the 00-component of the field equations
\begin{equation}
\frac{3}{c^{2}}\frac{d^{2}R}{dt^{2}}-\Lambda R=-\frac{4\pi G}{c^{4}}(\rho c^{2}+3P)R
\end{equation}
and for the space-components

\begin{equation}
\frac{R}{c^{2}}\frac{d^{2}R}{dt^{2}}+\frac{2}{c^{2}}\left(\frac{dR}{dt}\right)^{2}
+2\epsilon-\Lambda R^{2}=\frac{4\pi G}{c^{4}}(\rho c^{2}-P)R^{2}.
\end{equation}
From (5) and (6) one obtains

\begin{equation}
\frac{1}{c}\frac{d\rho}{dt}=-\frac{3}{Rc}\frac{dR}{dt}\left(\rho+
\frac{P}{c^{2}}\right).
\end{equation}
Using the approximations $P=0$ and $P=\frac{\rho c^{2}}{3}$ for the 
"non-relativistic approximation of the present universe" and a "radiation-dominated
universe", respectively, then from (7) one obtains

\begin{equation}
\rho_{mat}(t)R(t)^{3}=:A_{mat}\quad\mbox{for}\;P=0,\quad
\rho_{str}(t)R(t)^{4}=:A_{str}\quad\mbox{for}\; P=\frac{\rho c^{2}}{3},
\end{equation}
where $\rho_{mat},\rho_{str}$ denote mass and radiation density, respectively
and where $A_{mat}$ and $A_{str}$ are constants. Assuming that there is no coupling
between radiation and matter one puts
\begin{equation}
\rho:=\rho_{mat}+\rho_{str}.
\end{equation}
Then from (5) and (6) one obtains the Friedmann equation

\begin{equation}
\left(\frac{dR}{dt}\right)^{2}=\frac{\alpha}{R}+\frac{\sigma}{R^{2}}+
\frac{1}{3}\Lambda c^{2}R^{2}-\epsilon c^{2},
\end{equation}
where
\begin{equation}
\alpha:=\frac{8\pi G}{3}A_{mat},\quad \sigma:=\frac{8\pi G}{3}A_{str},\quad
\alpha\geq 0,\sigma\geq 0, \epsilon=0,\pm 1.
\end{equation}
Further restrictions of the parameters $\Lambda,\epsilon,\alpha,\sigma$ are not
used at the moment.

$\Lambda$ is considered as a fundamental natural constant connected with the basic
structure of space-time. Since the time of the appearance of the Einstein-de Sitter
model 1932 (see [2]), where $\Lambda=0$ is assumed, the case $\Lambda=0$ was
generally preferred. However recently the case $\Lambda>0$ was again taken into
account because of new observations. Therefore the focus of this note is directed to
this case. First a brief overview on the case $\Lambda=0$ is given.

\section{The case $\Lambda=0$}

In this case the Friedmann equation (10) reads
\begin{equation}
\left(\frac{dR}{dt}\right)^{2}=\frac{\alpha}{R}+\frac{\sigma}{R^{2}}-\epsilon c^{2},
\quad \epsilon=0,\pm 1
\end{equation}

\subsection{The case $\epsilon=0,-1,\,\alpha+\sigma>0$}

We put $\epsilon=-\delta,\,\delta=0,1$. Then
\[
\alpha R+\sigma+\delta c^{2}R^{2}>0,\quad R>0,
\]
i.e. there is no restriction for $R$ w.r.t. possible solutions. In particular,
there is no constant (stationary) solution $R(t)=const$. Either always
$R'(t)>0$ or $R'(t)<0$. In the first case $R(\cdot)$ is strongly monotonically
increasing and it turns out that it increases asymptotically for $t\rightarrow
\infty$ like $const\cdot t$ for $\epsilon=-1$ and like $const\cdot t^{2/3}$
for $\epsilon=0$. The second case $R'(t)<0$ can be excluded.

\vspace{3mm}

The case $\epsilon=0$ is called the {\em Einstein-de Sitter -model}. If one
introduces the so-called Hubble-parameter
\[
H(t):=\frac{R'(t)}{R(t)}
\]
then, using (11),(8),(9), the equation (12) can be written in the form
\[
\frac{3H(t)^{2}}{8\pi G}=\rho(t).
\]
Since the present value $H(t_{0})$ is approximatively known, one can define
a so-called critical density
\[
\rho_{cr}:=\frac{3H(t_{0})^{2}}{8\pi G}.
\]

\subsection{The case $\epsilon=+1,\,\alpha+\sigma>0$}

In this case (10) reads
\[
\left(\frac{dR}{dt}\right)^{2}=\frac{1}{R^{2}}(\alpha R+\sigma-c^{2}R^{2}).
\]
If $\alpha=0$ then there is an exceptional radius $R_{0}:=\frac{\sqrt{\sigma}}
{c}$. The region $R>R_{0}$ is a {\em forbidden region} for solutions because 
in this case $\sigma-c^{2}R^{2}<0$. The straight line $R(t)=R_{0}$ represents
an unstable stationary solution. The region $0<R<R_{0}$ is allowed. There are two
different types of solutions, the first one with $R'(t)>0$ starts at $R=0$ and
finished at $R=R_{0}$, the second one reversal. That is, each point of the constant
solution $R(t)=R_{0}$ is a branching point for solutions, reflecting the
unstability of this stationary solution.

If $\alpha>0$, then again there is an exceptional radius,
\[
R_{0}:=\frac{\alpha}{2c^{2}}\left(1+\sqrt{1+\frac{2\sigma c^{2}}{\alpha}}\right),
\]
with the same implications as before.

\subsection{Criterion for the value of the curvature parameter $\epsilon$}

Writing the equation (12) in the form
\begin{equation}
H(t)^{2}=\frac{8\pi G}{3}\rho(t)-\frac{\epsilon c^{2}}{R^{2}},
\end{equation}
the following criterion is obvious:
\begin{equation}
\epsilon\;\mbox{equals +1,-1,0 iff}\;
\rho(t_{0})>\rho_{cr}\;,\rho(t_{0})<\rho_{cr}\;,\rho(t_{0})=\rho_{cr}\;,
\mbox{respectively}.
\end{equation}

\section{The case $\Lambda>0$}

Obviously, there is a large distinction between the cases $\Lambda=0$ and
$\Lambda>0$ because $\Lambda$ is the coefficient of the "highest term" in
the right hand side of (10). Therefore it would be difficult to start with a
perturbation theoretic ansatz. For convenience we put
\[
\beta:=\frac{1}{3}\Lambda c^{2},\quad \gamma:=\epsilon c^{2},
\]
such that (10) now reads
\begin{equation}
\left(\frac{dR}{dt}\right)^{2}=\frac{\beta}{R^{2}}\left(R^{4}-\frac{\gamma}{\beta}
R^{2}+\frac{\alpha}{\beta}R+\frac{\sigma}{\beta}\right).
\end{equation}
We put
\[
p(R):=R^{4}-\frac{\gamma}{\beta}R^{2}+\frac{\alpha}{\beta}R+\frac{\sigma}{\beta}.
\]
That is
\[
p(R)=Rq(R)+\frac{\alpha}{\beta}, 
\]
where
\[
q(R)=R^{3}-\frac{\gamma}{\beta}R+\frac{\alpha}{\beta}.
\]
As already mentioned, $\Lambda$ is considered as an so far unknown fundamental 
constant
and $\epsilon$ is also a fixed (but also a priori unknown) constant. Thus the 
solutions of (15) depend on $\alpha$ and $\sigma$. Obviously a classification of them
depends strongly on the properties of the polynomial $p$ of the fourth degree.
Note that the "shape" of $p$ is independent of $\sigma$. This fact suggests 
to use the polynomial $q$ and the parameter $\sigma$ as preliminary parameters for
the classification.

The strategy for the classification is to start with the discriminant $D$ 
of $q$, given by
\begin{equation}
D:=\left(\frac{\alpha}{2\beta}\right)^{2}-\left(\frac{\gamma}{3\beta}\right)^{3}
=\frac{27\alpha^{2}\beta-4\gamma^{3}}{4\cdot 27\cdot \beta^{3}}.
\end{equation}
The parameter $D$ is decisive for the properties of the roots of $q$.
Therefore, first we recall the following facts
from the theory of zeros of polynomials of the third degree, applied to $q$.
\begin{itemize}
\item[(i)]
If $D<0$ then $q$ has two different real simple roots $0\leq R_{1}<R_{2}$.
Note that $\alpha>0$ implies $R_{1}>0$, but for $\alpha=0$ one obtains
$R_{1}=0$.
\item[(ii)]
If $D=0$ then $q$ has a positive double-root $R_{1}=R_{2}$.
\item[(iii)]
If $D>0$ then $q$ has no non-negative real root.
\end{itemize}
Note that (16) implies that in the cases $\epsilon=0$ and $\epsilon=-1$
always $D>0$ follows. This means that for these cases always the case (iii)
is true. This implies that also the polynomial $p$ has no positive real roots
for all $\sigma\geq 0$. This suggests first to separate this special case.

\subsection{The case $\epsilon=0,-1$ and $\sigma\geq 0$}

In this case $D>0$ follows and consequently there are no restrictions for 
$R>0$, i.e. there are no forbidden regions or separating straight lines.

\subsection{The case $\epsilon=+1$ and $\sigma=0$}

In the case $\epsilon=+1$ one has to take into account all three possibilities 
for $D$. This suggests to start with the case $\sigma=0$.

The roots $R_{1},R_{2}$ in the case (i)
can be calculated explicitly. The result is
\begin{equation}
R_{1}:=2\left(\frac{\gamma}{3\beta}\right)^{1/2}\cos\left(\frac{2\pi}{3}-
\frac{\phi}{3}\right),\quad R_{2}:=2\left(\frac{\gamma}{3\beta}\right)^{1/2}
\cos\frac{\phi}{3},
\end{equation}
where
\begin{equation}
\cos\phi=-\frac{\alpha}{2\beta}\left(\frac{\gamma}{3\beta}\right)^{-3/2}=
-\frac{1}{2}\sqrt{27}\gamma^{-3/2}\cdot\alpha\beta^{1/2},
\end{equation}
where
\[
-1<\cos\phi\leq 0 \quad \mbox{or}\quad \pi>\phi\geq\frac{\pi}{2}.
\]
The limit case $\phi=\pi/2$ corresponds to the case $\alpha=0$ and the limit case
$\phi=\pi$ to the case $D=0$, i.e. to the case of maximal $\alpha$. In the latter
case, where $R_{1}=R_{2}$, one has $\cos\frac{\pi}{3}=\frac{1}{2}$ such that in
this case
\[
R_{1}=R_{2}=\Lambda^{-1/2}.
\]
Note that in this case, where $\epsilon=+1$,  one has
\begin{equation}
A_{mat}=\frac{M}{2\pi^{2}},
\end{equation}
where $M$ is the total mass (dust) of the universe, because in this case the volume
of the 3-sphere is $2\pi^{2}R(t)^{3}$. Therefore,
\begin{equation}
\cos\phi=-\frac{2G}{\pi c^{2}}M\Lambda^{-1/2}
\end{equation}
and $D<0$ means
\[
M\Lambda^{1/2}<\frac{\pi c^{2}}{2G}.
\]
This means that in the case $\sigma=0,\,D<0,$ one has the following situation: The
interval $R\in (R_{1},R_{2})$ is a {\em forbidden region} for solutions because
in this case $p(R)<0$, i.e. there is an {\em upper region} $R>R_{2}$ and a
{\em lower region} $0<R<R_{1}$, where solutions are possible, except for the case
$\alpha=0$: in this case the lower region disappears. Moreover, in the case
$\alpha>0$ one obtains two different unstable stationary solutions
$R(t):=R_{1}$ and $R(t):=R_{2}$.

\vspace{3mm}

A lower region appeared already in the case $\epsilon=+1,\Lambda=0,\alpha>0$ (see
2.2). Recall that for $\sigma=0$ the exceptional radius is $R_{0}=\alpha c^{-2}$.
The new feature in the case $\Lambda>0$ consists in the appearance of a 
completely new upper region where solutions are possible. If we consider
the limit $\Lambda\rightarrow 0$ then $R_{2}\rightarrow\infty$ and
$R_{1}\rightarrow\alpha c^{-2}$. This follows from (17) and (18). One has
\[
\frac{\cos\left(\frac{2\pi}{3}-\frac{\phi}{3}\right)}{-\cos\phi}=
\frac{\gamma R_{1}}{3\alpha}.
\]
If $\beta\rightarrow 0$ then (18) yields $\phi\rightarrow\pi/2$. Now
\[
\lim_{\phi\rightarrow\pi/2}\frac{\cos\left(\frac{2\pi}{3}-\frac{\phi}{3}\right)}
{-\cos\phi}=\frac{1}{3},
\]
that is $\frac{1}{3}=\frac{\gamma}{3\alpha}\lim_{\beta\rightarrow 0}R_{1}$.

\vspace{3mm}

The special case $\alpha=0$, together with $\Lambda>0,\sigma=0$ as before,
is a limit case in so far as in this case the "new" upper region is present
but the "old" lower region disappeared. In this limit case (10) reads -
with $\epsilon=+1$
\[
\left(\frac{dR}{dt}\right)^{2}=\beta R^{2}-c^{2}.
\]
The straight line 
\[
R_{0}:=\frac{c}{\sqrt{\beta}}=\sqrt{\frac{3}{\Lambda}}
\]
separates the upper region
$R>R_{0}$ from the forbidden region $R<R_{0}$. Again one has the unstable
stationary solution $R(t):=R_{0}$. The solutions in the upper region represent
expanding universes where for $t\rightarrow\infty$ the solution $R(t)$
is asymptotically given by 
\[
\mbox{exp}\left({\sqrt{\frac{\Lambda}{3}}}\right). 
\]
These solutions are due to
W. de Sitter (see [3,\,4]). They are called {\em de Sitter universes}.

\vspace{3mm}

In the case $\sigma=0,\,D=0$ the forbidden region degenerates to a straight line,
i.e. the point $R_{1}=R_{2}$ is a 3-fold branching point for solutions because
an initial point $\{t_{0},\,R_{1}=R_{2}\}$ has three possibilities to
evolve: to remain stationary, to enter the upper or the lower region.

\vspace{3mm}

If $\sigma=0$ and $D>0$ then there is no restriction for $R>0$. This means that the
solutions can be divided into two disjoint classes, defined by $R'(t)>0$
or $R'(t)<0$ for all $t$. The latter class can be excluded.

\subsection{The case $\epsilon=+1$ and $\sigma>0$}

First let $D<0$. Then one obtains immediately from the structure of $p$ that
if $\sigma$ increases then the forbidden region decreases, i.e.
\[
\sigma_{1}<\sigma_{2}\quad\mbox{implies}\quad R_{1}(\sigma_{1})<R_{1}(\sigma_{2})<
R_{2}(\sigma_{2})<R_{2}(\sigma_{1})
\]
This means: There is a critical value $\sigma_{cr}$ such that
\[
R_{1}(\sigma_{cr})=R_{2}(\sigma_{cr})=:R_{cr}.
\]
The critical radius $R_{cr}$ can be calculated explicitly. This term depends
only on the derivation of the polynomial
\[
\frac{d}{dR}\left(Rq(R)\right)=4\left(R^{3}-\frac{\gamma}{2\beta}R+\frac{\alpha}
{4\beta}\right).
\]
Again this polynomial has two different positive real roots
$0<R'_{1}<R'_{2}$ which realize the maximum and the minimum of the polynomial
Rq(R), respectively. Further $R_{cr}=R'_{2}$. The calculation of $R_{cr}$ yields
\begin{equation}
R_{cr}=2\left(\frac{\gamma}{6\beta}\right)^{1/2}\cos\frac{\psi}{3}=
\left(\frac{2}{\Lambda}\right)^{1/2}\cos\frac{\psi}{3},
\end{equation}
where $\cos\psi=\frac{1}{\sqrt{2}}\cos\phi$, i.e. $\frac{3\pi}{4}>\psi\geq\frac
{\pi}{2}$. This means, in the massless case one has $\psi=\frac{\pi}{2}$, as 
before, and in the limit case of maximal mass $\psi=\frac{3\pi}{4}$, or
$\psi/3=\pi/4$, which implies that in the limit case one has
\[
R_{1}=R_{2}=R_{cr}=\frac{1}{\sqrt{\Lambda}}.
\]
Then a simple calculation yields
\[
R_{cr}q(R_{cr})=-\frac{2}{3}\frac{\gamma^{2}}{\beta^{2}}\cos^{2}\frac{\psi}{3}
\cdot\cos\frac{2\psi}{3},
\]
and the critical $\sigma$-value is $\sigma_{cr}:=-\beta R_{cr}q(R_{cr})$, i.e.
\begin{equation}
\sigma_{cr}=\frac{2c^{2}}{\Lambda}\cos^{2}\frac{\psi}{3}\cdot\cos\frac{2\psi}{3}.
\end{equation}
That is, if $D<0$ then one has to distinguish between three cases:
\begin{itemize}
\item[(i)]
$\sigma<\sigma_{cr}$: Then the interval $(R_{1}(\sigma),R_{2}(\sigma))$ is a forbidden 
region,
\item[(ii)]
$\sigma=\sigma_{cr}$: The forbidden region degenerates to a straight line,
\item[(iii)]
$\sigma>\sigma_{cr}$: There is no restriction for $R>0$.
\end{itemize}

\section{Summary}

The foregoing discussion of several cases can be summarized as follows. For simplicity
the branching cases $D=0,\,\sigma=0$ and $D<0,\,\sigma=\sigma_{cr}$ are 
omitted, also the case $\alpha=0$ and the cases where there is no time interval
with $R'(t)>0$. There are four distinct types of solutions:
\begin{itemize}
\item[(A)]
$\epsilon=0$ or $\epsilon=-1$. Further $\sigma\geq 0$. (Recall $D>0$ in this case.)
\item[(B)]
$\epsilon=+1$ and: $D>0,\,\sigma\geq 0$ or $D<0,\,\sigma>\sigma_{cr}$.
\item[(C)]
$\epsilon=+1$ and: $D<0,\,\sigma<\sigma_{cr}$. (Solution in the upper region.)
\item[(D)]
$\epsilon=+1$ and: $D<0,\,\sigma<\sigma_{cr}$. (Solution in the lower region.)
\end{itemize}
Next as the counterpart the essential analytic-geometric characteristic
properties of the cases (A)-(D) are described. We choose $t_{0}=0$ as the
starting time.
\begin{itemize}
\item[(A)]
$R(0)=0$, for $t\rightarrow\infty$ the radius (scalar-factor) $R(t)$ behaves
approximatively like $\mbox{exp}(\sqrt{\beta}t)$, there is a turning point
$t_{w}$ for $R(\cdot)$; $H(t)>0$ is strongly monotonically decreasing.
\item[(B)]
$R(0)=0$, for $t\rightarrow\infty$ the radius $R(t)$ behaves approximatively
like $\mbox{exp}(\sqrt{\beta}t)$, there is a turning point $t_{w}$
for $R(\cdot)$; $H(t)>0$,
there is a minimum point $t_{min}$ for $H(\cdot),\; H(t)>H(t_{min})$ for
$t>t_{min}$ and there is a turning point $t_{w,H}>t_{min}$ for $H(\cdot)$.
\item[(C)]
$R(0)=R_{2}(\sigma)$, for $t\rightarrow\infty$ the radius $R(t)$ behaves
approximatively like $\mbox{exp}(\sqrt{\beta}t),\; R'(0)=0,\; H(t)>0$ is
strongly monotonically increasing.
\item[(D)]
$R(0)=0$, there is a maximum point $t_{max}$ for $R(\cdot),\; R(t_{max})=
R_{1}(\sigma)$; $H(t)>0$ is strongly monotonically decreasing.
\end{itemize}

\section{Conclusions}

A celebrated observation on the large-scale structure of the universe was the
discovery that the universe expands, the scale-factor $R$ increases. This is
an implication of the observation that the Hubble-constant (the Hubble-parameter
of the presence) is positive.

Recent observations on very far distant Supernovae, e.g. the {\em Supernovae
Cosmology Project} (SCP) or the observations by WMPA-satellites, show that the
expansion is even {\em accelerating} (see e.g. S. Perlmutter [5]).
This means that $H(\cdot)$ is neither
constant nor decreasing at present, i.e. in a former epoch $H$ was smaller than
to-day. This led - in contrast to the long accepted {\em Einstein-de Sitter-model}
(see [2]) - to the conclusion that one has to assume rather that $\Lambda>0$.
This assumption is compatible with the interpretation of the expansion
as an intrinsic property of "space", which leads consequently to an exponential
growth or increase of it (see e.g. Lemaitre [6]).

If these mentioned observations will be confirmed by further observation
projects, then case (A) could be excluded, i.e. then necessarily $\epsilon=+1$,
only a spherical model could be the right one. Also (D) could be excluded, i.e.
the so-called BigBang-BigCrunch solution.

On the contrary, the models (B) and (C) are compatible with the mentioned
observations.

The model (B) represents a BigBang solution, whereas the model (C) can be
considered as an Anti-BigBang solution. Interestingly enough, this model
is essentially the model of Lemaitre (see [6]), it does not have a singular
origin but it starts as a static Einstein-Universe (see e.g. [7]) and can
be described for $t\rightarrow\infty$ asymptotically by the de Sitter-Universe
(see [3,4], see also [8]). 

That is, these observations lead to the alternative between the BigBang-model (B)
and the Anti-BigBang model (C). At present the BigBang-hypothesis is widely
accepted, which is supported by the 3K-background radiation and the distribution
of Helium and Hydrogen in the universe. However, still it remains a hypothesis.

If it would turn out by further observations in the future, for example that in
far former epochs the Hubble parameter $H$ was much larger than today, i.e. that
$H$ was decreasing in that time, then this would be a proof for the BigBang-model
(of course within the framework of the Friedmann equation), because in this case
$H$ must have passed through a minimum and only in the model (B) there appears such
a minimum for $H$.

According to the Summary the alternative between the models (B) and (C) can
be expressed by their characteristic parameters as
\[
D>0, \mbox{or}\; D<0\; \mbox{and}\; \sigma>\sigma_{cr}\quad \mbox{versus}\quad
D<0\; \mbox{and}\; \sigma<\sigma_{cr}.
\]
Finally it should be mentioned that also in the case $\Lambda>0$ the Friedmann
equation (10) can be formally written in the form (13) if one puts
\[
\rho(t):=\rho_{mat}(t)+\rho_{str}(t) +\frac{\Lambda c^{2}}{8\pi G}.
\]
Then again the alternative (14) remains true. However, since all models 
with $\epsilon =0,-1$ are now excluded one can state
\begin{equation}
\rho(t_{0})>\rho_{cr}=\frac{3H(t)^{2}}{8\pi G},\quad t_{0}\; \mbox{present time}.
\end{equation}
According the foregoing analysis (23) does {\em not} imply that the
BigBang-BigCrunch model (D) is true.

\section{ The special case $D>0,\;\sigma=0$}

In this special (B)-model it is easy to calculate the characteristic points
$R_{w}=R(t_{w}), R_{min}=R(t_{min}), H(t_{min})^{2}, R_{w,H}=R(t_{w,H})$ and
$H(t_{w,H})^{2}$ (see [9]). First note that
\[
R_{w}<R_{min}<R_{w,H}
\]
because of $D>0$. Using the parameters $M,\Lambda,c,G$ one obtains
\[
D>0\quad \mbox{means}\quad \Lambda>\left(\frac{\pi}{2}\frac{c^{2}}{GM}\right)^{2}.
\]
Further
\[
R_{w}=\left(\Lambda\cdot\frac{\pi}{2}\frac{c^{2}}{GM}\right)^{-1/3},
R_{min}=\frac{2}{\pi}\frac{GM}{c^{2}},\quad R_{w,H}=\frac{3}{\pi}\frac{GM}{c^{2}},
\]
\[
H(t_{min})^{2}=\frac{1}{3}c^{2}\left(\Lambda-\left(\frac{\pi}{2}\frac
{c^{2}}{GM}\right)^{2}\right),
H(t_{w,H})^{2}=\frac{1}{3}\left(\Lambda-\frac{20}{27}\left(\frac{\pi}{2}
\frac{c^{2}}{GM}\right)^{2}\right).
\]

\section{Some calculations for the Anti-BigBang model}

In this case one has $D<0$ and $\sigma<\sigma_{cr}$. We are interested in the case of 
nearly maximal mass $M$. According to (20) one has
\[
M\Lambda^{1/2}=\frac{\pi c^{2}}{2G}(-\cos\phi),\quad \frac{\pi}{2}\leq\phi<\pi.
\]
For convenience we put
\[
\frac{2\psi}{3}:=\frac{\pi}{2}-\chi,\quad \chi>0,\quad\chi\;\mbox{small}.
\]
Then $\frac{\psi}{3}=\frac{\pi}{4}-\frac{\chi}{2}$ and $\psi=
\frac{3\pi}{4}-\frac{3\chi}{2}$ hence
$\cos\psi=\frac{1}{\sqrt{2}}(-\cos\frac{3\chi}{2}+\sin\frac{3\chi}{2})$.
Further one has $\cos\phi=\sqrt{2}\cos\psi$, i.e.
\[
M\Lambda^{1/2}=\frac{\pi c^{2}}{2G}(\cos\frac{3\chi}{2}-\sin\frac{3\chi}{2}).
\]
Since $\chi$ is small, in the first approximation w.r.t. $\chi$, i.e. neglecting
higher terms than $\chi$, one has
\begin{equation}
M\Lambda^{1/2}=\frac{\pi c^{2}}{2G}(1-\frac{3\chi}{2}).
\end{equation}
Further one has $\cos(\frac{\pi}{4}-\frac{\chi}{2})=\frac{1}{\sqrt{2}}
(\cos\frac{\chi}{2}+\sin\frac{\chi}{2})$, hence
$\cos^{2}(\frac{\pi}{4}-\frac{\chi}{2})=\frac{1}{2}(1-\sin\chi)$ and
\[
\cos^{2}\frac{\psi}{3}\cdot\cos\frac{2\psi}{3}=\frac{1}{2}\sin\chi(1+\sin\chi)
=\frac{1}{2}\cdot\chi
\]
follows in the first approximation. Thus one obtains
\begin{equation}
\sigma_{cr}=\frac{c^{2}}{\Lambda}\cdot\chi.
\end{equation}
The result is: If (24) is satisfied where $\chi>0$ is small and $\sigma<c^{2}
\Lambda^{-1}\chi$, then the corresponding Friedmann-model is an Anti-BigBang model
(C). 

Finally, we calculate the minimal radius in this case for the limit case
$\sigma=\sigma_{cr}$. According to (21) one gets
\[
R_{cr}=\left(\frac{2}{\Lambda}\right)^{1/2}\cos\frac{\psi}{3}=
\left(\frac{2}{\Lambda}\right)^{1/2}\cos(\frac{\pi}{4}-\chi)=
\Lambda^{-1/2}(\cos\frac{\chi}{2}+\sin\frac{\chi}{2}),
\]
i.e. in the first order approximation one has
\[
R_{cr}=\frac{1}{\sqrt{\Lambda}}\left(1+\frac{\chi}{2}\right).
\]
Using (11) one obtains from (25)
\[
A_{str,cr}=\frac{3}{4\pi^{2}}\left(\frac{\pi c^{2}}{2G}\right)\cdot\frac{\chi}
{\Lambda}.
\]
In the limit case $D=0$, i.e. $\chi=0$, this means
\[
M\Lambda^{1/2}=\frac{\pi c^{2}}{2G}
\]
and in this case one obtains
\[
R_{cr}=\left(\frac{\pi c^{2}}{2G}\right)^{-1}\cdot M.
\]
The constant $\mu:=\frac{\pi c^{2}}{2G}$ is connected with $\kappa$ (cf.
equation (1)) by $\mu\cdot\kappa=4\pi^{2}$.

\section{References}

\begin{enumerate}
\item
11.Vorlesung Kosmologie: page.mi.fu-berlin.de/sfroehli/RelTheorie/kapitel 11.pdf\\
\item
A. Einstein and W. de Sitter: Proc. Nat. Acad. of Science 18, 213 (1932)\\
\item
W. de Sitter: On the relativity of inertia: Remarks concerning Einsteins
latest hypothesis, Proc. Kon. Ned. Acad. Wet. 19, 1217-1225 (1917)\\
\item
W. de Sitter: On the curvature of space, Proc. Kon. Ned. Acad. Wet. 20, 229-243
(1917)\\
\item
S. Perlmutter: Supernovae, Dark Energy and the Accelerating Universe,
Physics Today, April 2003, 53-60\\
\item
G. Lemaitre: Annales de la Societe Scientifique de Bruxelles, A 47, 49 (1927)\\
\item
A. Einstein: Sitzungsber. Preuss. Akad. Wiss., 142 (1917)\\
\item
S. R\"ohle: Mathematische Probleme in der Einstein-de Sitter Kontroverse,
U Mainz, FB Math. Preprint MPI Berlin, pdf
\item
H. Baumg\"artel: Friedmann equation and SCP-results, arXiv: 1106.3258v1 [math-ph]
(2011)
\end{enumerate}

\end{document}